\documentclass[
reprint,
superscriptaddress,
amsmath,
amssymb,
showkeys,
aps,
prb,
nolongbibliography,
]{revtex4-2}

\usepackage{threeparttable}
\usepackage{setspace}
\usepackage{makecell}
\usepackage{amsmath}
\usepackage{graphicx}
\usepackage{dcolumn}
\usepackage{bm}
\usepackage{hyperref}
\usepackage[mathlines, pagewise]{lineno}
\usepackage{textcomp}
\usepackage{gensymb}
\usepackage{subfigure}
\usepackage{url}
\usepackage{xcolor}
\usepackage{soul}
\usepackage{hhline}
\usepackage[version=3]{mhchem}
\usepackage{multirow}
\usepackage[american]{babel}
\usepackage{textgreek}
\usepackage{miller}

\begin{document}

\title{
Self-assembling of multilayered polymorphs with ion beams
}

\author{Alexander Azarov}
\affiliation{Department of Physics and Centre for Materials Science and Nanotechnology, University of Oslo, PO Box 1048 Blindern, N-0316 Oslo, Norway}

\author{Cristian Radu}
\affiliation{National Institute of Materials Physics, Atomistilor 405A, 077125 Magurele, Romania}

\author{Augustinas Galeckas}
\affiliation{Department of Physics and Centre for Materials Science and Nanotechnology, University of Oslo, PO Box 1048 Blindern, N-0316 Oslo, Norway}

\author{Ionel Florinel Mercioniu}
\affiliation{National Institute of Materials Physics, Atomistilor 405A, 077125 Magurele, Romania}

\author{Adrian Cernescu}
\affiliation{Attocube systems AG, Eglfinger Weg 2, 85540 Haar, Germany}

\author{Vishnukanthan Venkatachalapathy}
\affiliation{Department of Physics and Centre for Materials Science and Nanotechnology, University of Oslo, PO Box 1048 Blindern, N-0316 Oslo, Norway}

\author{Edouard Monakhov}
\affiliation{Department of Physics and Centre for Materials Science and Nanotechnology, University of Oslo, PO Box 1048 Blindern, N-0316 Oslo, Norway}

\author{Flyura Djurabekova} 
\affiliation{Department of Physics and Helsinki Institute of Physics, University of Helsinki, P.O. Box 43, FI-00014, Finland}

\author{Corneliu Ghica} \email{cghica@infim.ro}
\affiliation{National Institute of Materials Physics, Atomistilor 405A, 077125 Magurele, Romania}

\author{Junlei Zhao} \email{zhaojl@sustech.edu.cn}
\affiliation{Department of Electrical and Electronic Engineering, Southern University of Science and Technology, Shenzhen 518055, China}

\author{Andrej Kuznetsov} \email{andrej.kuznetsov@fys.uio.no}
\affiliation{Department of Physics and Centre for Materials Science and Nanotechnology, University of Oslo, PO Box 1048 Blindern, N-0316 Oslo, Norway}

\begin{abstract} 

Polymorphism contributes to the diversity of nature, so that even materials having identical chemical compositions exhibit variations in properties because of different lattice symmetries. 
Thus, if stacked together into multilayers, polymorphs may work as an alternative approach to the sequential deposition of layers with different chemical compositions. 
However, selective polymorph crystallization during conventional thin film synthesis is not trivial; e.g. opting for step-like changes of temperature and/or pressure correlated with switching from one polymorph to another during synthesis is tricky, since it may cause degradation of the structural quality. 
In the present work, applying the disorder-induced ordering approach we fabricated such multilayered polymorph structures using ion beams. 
We show that during ion irradiation of gallium oxide, the dynamic annealing of disorder may be tuned towards self-assembling of several polymorph interfaces, consistently with theoretical modelling. 
Specifically, we demonstrated multilayers with two polymorph interface repetitions obtained in one ion beam assisted fabrication step. 
Importantly, single crystal structure of the polymorphs was maintained in between interfaces exhibiting repeatable crystallographic relationships, correlating with optical cross-sectional maps. 
This data paves the way for enhancing functionalities in materials with not previously thought capabilities of ion beam technology.

\end{abstract}

\maketitle

Combination of materials into multilayers is a prominent strategy for enhancing properties of solids. 
Conventionally such structures are synthesized during thin film deposition by changing chemical composition of the layers [1]. 
However, in materials exhibiting polymorphism such structures can be realized without compositional changes, by stacking one polymorph on another in a controllable way. 
Indeed, since different polymorphs crystallize in different lattices and, consequently, exhibit variations of physical properties, polymorphism can be used for extending material functionalities [2]. 
For example, polymorphic heteroepitaxy have been developed for transition-metal dichalcogenides, where metallic and semiconducting \ce{MoTe2} polymorphs were integrated into a single heterostructure [3]. 
Gallium oxide (\ce{Ga2O3}) is another example of solids having multiple polymorphs. 
These polymorphs exhibit different symmetries, with the monoclinic $\beta$-phase known to be the most stable under normal conditions [4]. 
Metastable \ce{Ga2O3} polymorphs, e.g. corundum-like $\alpha$-phase or orthorhombic $\kappa$-phase, were also synthesized in thin film forms applying variations in temperature/pressure [5]. 
However, since stacking different \ce{Ga2O3} polymorphs together might require temperature/pressure adjustments during the deposition process, it may jeopardize the integrity of the polymorph interfaces. 
Likely because of that reason, there are no data in literature reporting multilayered \ce{Ga2O3} polymorph structures having acceptable in applications quality of interfaces synthesized by conventional methods [6]. 
However, very recently, it has been shown that cubic spinel $\gamma$-\ce{Ga2O3} polymorph can be formed on the top of a $\beta$-phase substrate by disorder-induced ordering [7-11]. 
This fabrication was demonstrated at room temperature (RT) and the rationale is that the $\beta$-to-$\gamma$ transition occurs upon reaching a certain disorder threshold in the system [12], resulting in rather abrupt $\beta$/$\gamma$ interface [10]. 
Meanwhile, variations of the irradiation temperature and flux, strongly affecting the evolution of the primarily generated ``ballistic" disorder, may provide unprecedented self-organization opportunities in such structures via dynamic defect annealing [13,14]. 

\begin{figure*}[ht!]
 \includegraphics[width=14cm]{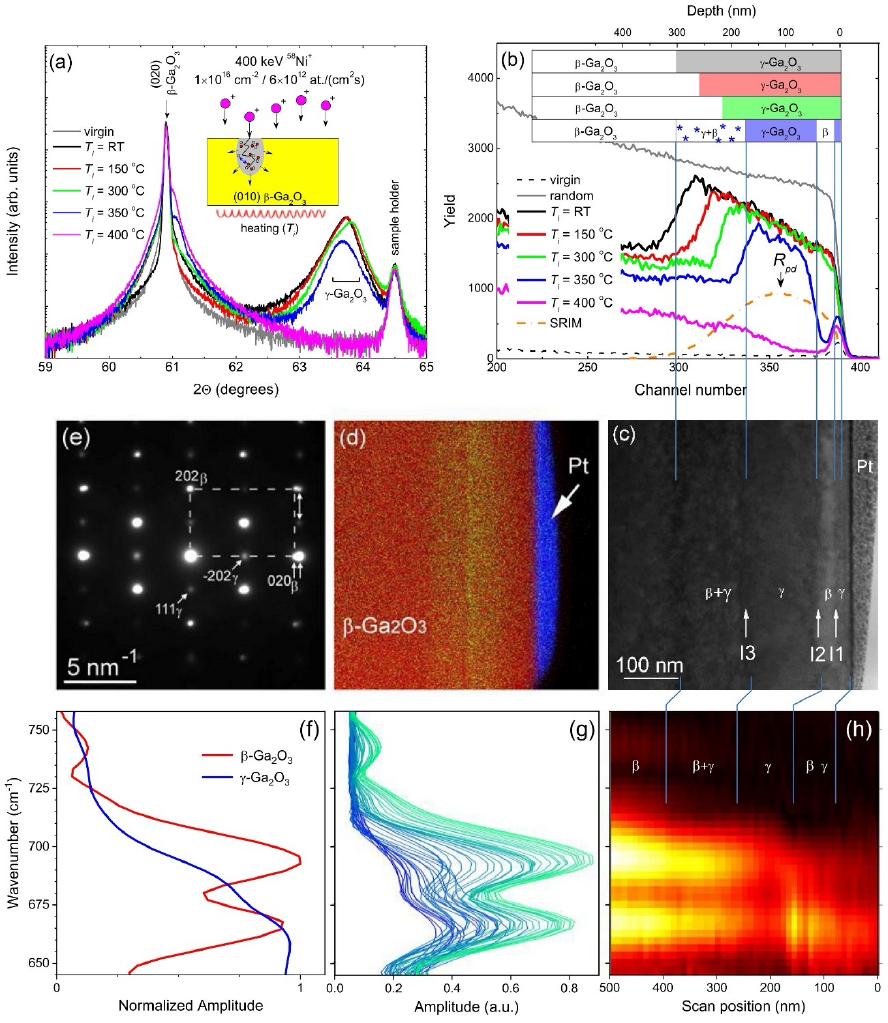}
 \caption{
    \textbf{Impact of the irradiation temperature on the disorder-induced $\beta$-to-$\gamma$ \ce{Ga2O3} phase transformation, including structural and optical map.} 
    (a) XRD 2$\Theta$ scans together with the cartoon in the inset representing collision cascade occurring in the samples; 
    (b) corresponding RBS/C spectra, including a hypothetical cross-section of the sample in the inset and the dashed-dotted line illustrating the nuclear energy loss vs depth profile calculated with SRIM code [21] simulations for the implanted ions and plotted in arbitrary units; 
    c) bright-field cross-section TEM image of the sample implanted at 350 $\degree$C with the interfaces aligned to the corresponding RBS/C data; 
    (d) Cross-sectional chemical elemental mapping of the lamella used for TEM investigations; 
    (e) SAED pattern taken from the whole irradiated part of the sample. 
    (f) nano-FTIR signatures of $\beta$ and $\gamma$ polymorphs recorded in plan-view and represented by the normalized absorption spectra, 
    (g) depth-resolved absorption spectra obtained by cross-sectional nano-FTIR scan from the surface down to 500 nm with 10 nm steps; 
    (h) nano-FTIR absorption spectra map revealing successive distribution of $\gamma$-$\beta$-$\gamma$-$\beta$ polymorphs in the cross-section aligned with the TEM and RBS/C data. 
    Notably, the identification of phases, as labeled in the panels (c) and (h) as well as in the cartoon in panel (b), is based on the HRTEM results as discussed latter in the paper.
 }
 \label{fig:1}
\end{figure*}

Therefore, in the present paper, we study the impact of the dynamic defect annealing on the disorder-induced phase transitions in $\beta$-\ce{Ga2O3}. 
Specifically, performing ion bombardment in a wide range of sample temperatures and ion fluxes, we found that the system can be indeed tuned towards self-assembling of several polymorph interfaces. 
As a result, we demonstrated the fabrication of multilayered $\gamma$/$\beta$ structures having pure $\gamma$-, $\beta$- or $\gamma$/$\beta$ mixture \ce{Ga2O3} phases separated in different layers; correlated with corresponding repetitions of optical properties. 
Thus, in addition to the fundamental novelty, these data pave the way for explorations of the \ce{Ga2O3} polymorph heterostructures to become a part of technology [15].

Fig. 1 sums up the data measured by (a) XRD, (b) RBS/C, (c-e) TEM, and (f-h) nano-FTIR – illustrating the impact of the irradiation temperature on the $\beta$-to-$\gamma$ \ce{Ga2O3} transformation, in correlation with variations of optical properties. 
Indeed, already XRD data alone suggest remarkable phase alterations, as monitored by the evolution of the peak at $\sim 63.75\degree$, assigned to $\gamma$-\ce{Ga2O3} consistently with literature [7-11], see Fig.~1(a). 
Spectacularly, this $\gamma$-\ce{Ga2O3} related peak, emerging upon RT irradiation and maintained upon a gradual temperature increase from RT to 350 $\degree$C, disappears in the sample irradiated at 400 $\degree$C; indicating that $\beta$-to-$\gamma$ phase transformation is challenged by the dynamic defect annealing. 
This is also supported by the RBS/C data in Fig.~1(b). 
Indeed, RT ion irradiation results in the formation of a box-like profile with a distinct channeling improvement towards the surface, known as a characteristic signature for the $\gamma$-\ce{Ga2O3} film formed on top of the $\beta$-\ce{Ga2O3} substrate [7,10]. 
The thickness of this box-like profile gradually shrinks towards the surface as a function of the irradiation temperature, e.g. for 150 $\degree$C and 300 $\degree$C irradiations, see the RBS/C spectra and the corresponding cartoons in the inset in Fig.~1(b). 
Meanwhile, the RBS/C data for the sample irradiated at 350 $\degree$C are of particular interest. 
Here, the shrinkage of the box-like profile – corresponding to the shrinkage of the $\gamma$-layer – occurs both from the bulk and from the surface, suggesting formation of several polymorph interfaces. 
Additionally, a low-intensity dechanneling shoulder is observed right behind the $\gamma$-layer signature in Fig.~1(b), as also illustrated with ``disordering'' symbols – crosses – in the inset. 
Concurrently, in the vicinity of the surface, there is a deviation from the $\beta$-like symmetry too, correlated with the enhancement of the corresponding RBS/C signal. 
Finally and consistently with XRD data in Fig.~1(a), the box-like $\gamma$-phase signature in the RBS/C spectra vanishes upon increasing irradiation temperature to 400 $\degree$C, see Fig.~1(b). 
Importantly, the ion flux is another critical parameter affecting the dynamic annealing processes [14], and corresponding data revealing the impact of the ion flux variations on the $\beta$-to-$\gamma$ phase transformation are included in Supplementary Note I.

\begin{figure*}[ht!]
 \includegraphics[width=14cm]{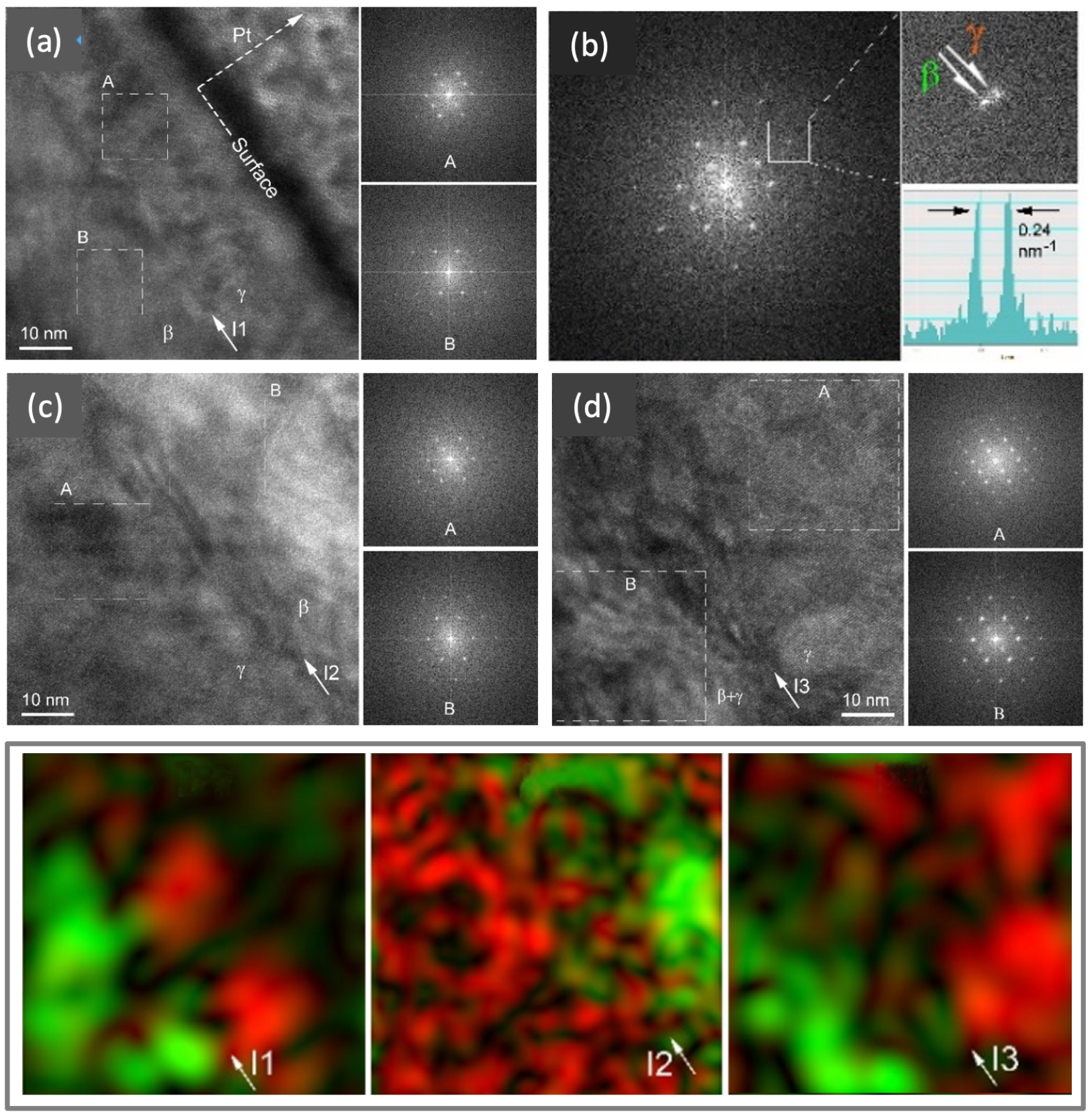}
 \caption{
    \textbf{Identification of polymorphs around interfaces in the multilayered structure.} 
    (a) HRTEM images and FFT patterns from areas labeled A and B on either side of the I1 interface verifying \ce{Ga2O3} phase separation; 
    (b) FFT pattern taken over both sides of I1 interface together with a magnified image illustrating the appearance of the extra spot adjacent to $020_{\beta}$ correlated with the presence of the $\gamma$-phase; 
    (c-d) HRTEM images and FFT patterns from areas labeled A and B around I2 and I3 interfaces, respectively; 
    (e) color maps around I1, I2 and I3 interfaces associated with 020β and -$404_{\gamma}$ spots in green and red, respectively. 
 }
 \label{fig:2}
\end{figure*}

Thus, XRD and RBS/C data in combination suggest the formation of a complex multilayered polymorph structure in the sample irradiated at 350 $\degree$C. 
To verify this conclusion, Fig.~1(c) shows the TEM cross-section of this sample, in direct correlation with the RBS/C depth scale. 
Indeed, Fig.~1(c) reveals several clearly defined layers and interfaces pointed by white arrows. 
These layers have thicknesses of 25, 30, 150 and 130 nm counting from the surface to the bulk, well-aligned with the RBS/C results shown in Fig.~1(b). 
Notably, the chemical analysis in Fig.~1(d) confirms the diluted state of the implanted impurity consistently with literature [10], see more data in Supplementary Note II. 
The selected-area electron diffraction (SAED) pattern in Fig.~1(e), acquired from the whole cross-section region in Fig.~1(c), is interpreted as an overlap of two lattice signatures having close lattice parameters. 
The dominating phase, considering the intensity of the diffraction spots, is assigned to the monoclinic $\beta$-\ce{Ga2O3} \hkl[10-1] zone axis orientation. 
Meanwhile, the weaker diffraction spots in intermediate positions with respect to the $\beta$-\ce{Ga2O3} pattern correspond to $\gamma$-\ce{Ga2O3} consistently with literature [16,17]. 

Equally importantly, distinctly different spectral nano-FTIR signatures of $\beta$- and $\gamma$-\ce{Ga2O3}, as systematically measured in planar geometry on control samples, see Fig.~1(f), enabled a meaningful nano-FTIR cross-sectional profiling of the sample irradiated at 350 $\degree$C – see Figs.~1(g) and (h). 
The correlation between structural cross-sections in Fig.~1(b-c) and optical map in Fig.~1(h) is striking – as indicated by vertical blue lines guiding eyes – even though the definition of the polymorph interface positions in the nano-FTIR map is less accurate, because of the tip size related uncertainties, as explained in Supplementary Note III.
Nevertheless, the data in Figs.~1(f-h) comprise probably the first ever demonstration of the functional properties modulation in polymorph multilayers fabricated with ion beams. 

Further insights into the physics of this amazing multilayered polymorph structures were obtained by HRTEM investigations (Fig.~2) around each of the interfaces as marked by arrows in Fig.~1(c). 
Specifically, Fig.~2(a) shows a HRTEM image together with the corresponding FFTs taken in the vicinity of I1 interface, confirming the separation of \ce{Ga2O3} into two layers, 25 nm and 30 nm thick, consistently with the lower magnification observation in Fig.~1(c). 
The features observed in the SAED patterns acquired from the whole multilayer structure in Fig.~1(e), are now examined in the FFT patterns from smaller areas, labelled as A and B on either side of the I1 interface, see Fig.~2(a). 
These FFT patterns show two different families of spots. 
The FFT pattern from B-area is indexed as the monoclinic $\beta$-\ce{Ga2O3} in \hkl[10-1] orientation. 
In its turn, the FFT pattern from A-area contains spots in nearly half-way positions with respect to that in B-area and indexed as the cubic defective-spinel $\gamma$-\ce{Ga2O3} following literature [16,17]. 
This FFT fingerprint change proves that \ce{Ga2O3} lattice has been locally reorganized into abruptly separated layers containing different polymorphs. 
Moreover, the FFT diagram taken from a bigger area across the I1 interface, see Fig.~2(b) and magnifications in the insets, validates that there are two adjacent characteristic spots clearly separated by $\Delta g/g \simeq 3\%$, corresponding to the peak splitting of 0.24 nm$^{-1}$.

Similar characterizations were carried out on areas around I2 and I3 interfaces as illustrated by Figs.~2(c) and (d), clearly defining regions with different crystal structures on either side of the interfaces. 
Indeed, I2 interface separates a layer of pure $\beta$-phase above ($\sim30$ nm thick) from a layer of pure $\gamma$-phase below ($\sim150$ nm thick) in excellent agreement with the interpretation of the RBS/C and nano-FTIR data in Figs.~1(b) and (h), respectively. 
In its turn, I3 interface separates the single-phase $\gamma$-layer above from a mixed $\beta$-$\gamma$ region deeper in the sample, containing predominantly $\beta$-phase with a minor portion of $\gamma$-phase. 
This is consistent with the RBS/C and nano-FTIR data, too. 
Complimentary, the phase color maps obtained with GPA software [18,19] around I1, I2, and I3 interfaces with $020_{\beta}$ and $-404_{\gamma}$ spots color-coded in green and red, visualize the polymorph flips across the interfaces, see in Fig.~2(e). 
More detailed structural information collected around each interface is available from the Supplementary Notes IV and V. 

\begin{figure*}[ht!]
 \includegraphics[width=14cm]{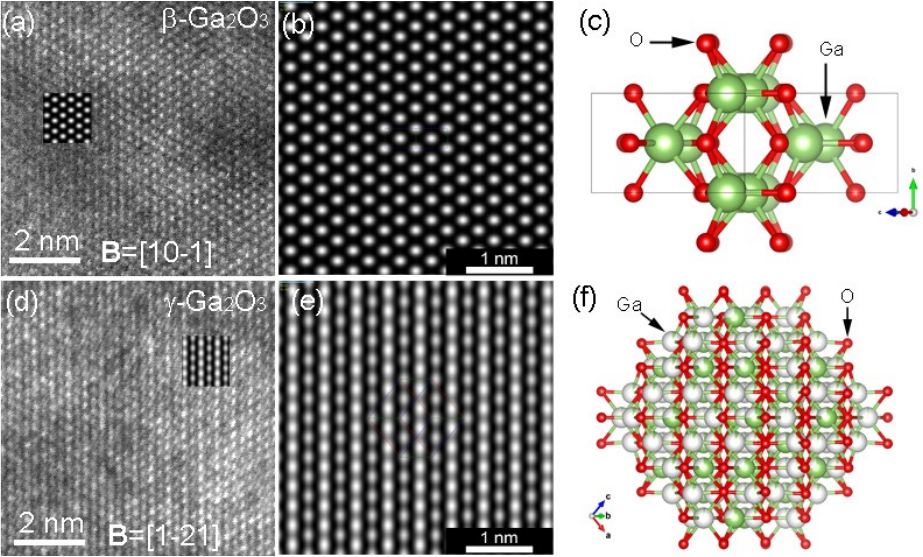}
 \caption{
    \textbf{Atomic-resolution $\beta$- and $\gamma$-phase images in comparison with simulations.} 
    (a) Experimental HRTEM image from a layer containing $\gamma$-\ce{Ga2O3} phase (simulated pattern inserted) together with (b) corresponding simulated HRTEM pattern and (c) structural model of the $\beta$-\ce{Ga2O3} unit cell in \hkl[10-1] orientation used in simulations. 
    (d) Experimental HRTEM image from a layer containing the $\gamma$-\ce{Ga2O3} phase (simulated pattern inserted) together with (e) the corresponding simulated HRTEM pattern and (f) structural model of the $\gamma$-\ce{Ga2O3} unit cell in \hkl[1-21] orientation used in simulations.
 }
 \label{fig:3}
\end{figure*}

Thus, the HRTEM analysis confirms a complex nature of the structural transformations in $\beta$-\ce{Ga2O3} upon ion bombardment at 350 $\degree$C, namely its self-assembling into a multilayered structure with alternating polymorph layers. 
Moreover, despite the contrast irregularities, likely because of the elastic strain induced by residual structural defects, the HRTEM data suggest that all parts of the sample conserved the single crystal structure, so that $\beta$ and $\gamma$ phases are well-aligned along the crystallographic relation defined by \hkl[10-1]$_{\beta}$ $\vert \vert$ \hkl[1-21]$_{\gamma}$, \hkl(-101)$_{\gamma}$ $\vert \vert$ \hkl(010)$_{\beta}$, and \hkl(111)$_{\gamma}$ $\vert \vert$ \hkl(101)$_{\beta}$ at the corresponding interfaces. 
Ultimately, Fig.~3 shows atomic-resolution analysis of the HRTEM data. Specifically, Figs.~3(a) and (d) show two typical HRTEM images cropped out of $\beta$ and $\gamma$ regions, respectively. 
The most striking feature is the double period of the fringes observed in the $\gamma$-phase with respect to that in the $\beta$-phase, correlating with the additional spots in halfway positions of the SAED patterns and distinctions in the FFT diagrams. 
The only consistent explanation of all these features is that we observe – indeed – repeatable separations of the monoclinic $\beta$-\ce{Ga2O3} from the cubic defective spinel $\gamma$-\ce{Ga2O3}. 
For comparison, we performed both electron diffraction and HRTEM image simulation using the ReciPro software [20] for both $\beta$ and $\gamma$ structures. 
The simulated images were generated in \hkl[10-1]$_{\beta}$ and \hkl[1-21]$_{\gamma}$ zone axes respectively, using the specimen thickness and the objective lens defocus as variable parameters (see Supplementary Note VI). 
Two representative simulated patterns best fitting the experimental images are shown in Figs.~3(b) and (e), as well as inserted – in form of smaller frames – into the corresponding experimental micrographs for direct comparison. 
The atomic models of the $\beta$-\ce{Ga2O3} unit cell in \hkl[10-1] orientation and $\gamma$-\ce{Ga2O3} unit cell in \hkl[1-21] orientation used in simulations are shown in Figs.~3(c) and (f), respectively. 
Comparing images in Fig.~3, we observe nearly doubling of periods between \hkl(202) planes in $\beta$-\ce{Ga2O3} ($d_{202\beta} = 0.236$ nm) versus \hkl(111) planes in $\gamma$-\ce{Ga2O3} ($d_{111\gamma} = 0.476$ nm). 

Thus, combining data in Figs.~1-3, we unambiguously show the formation of multilayered \ce{Ga2O3} polymorph structure via self-assembling governed by the dynamic annealing in the sample irradiated at 350 $\degree$C, also featuring corresponding repetitions of the optical signatures. 
At this end, a consistent scenario for such multilayered polymorph structure self-assembling may be the following. 
The $\gamma$-layer in between I3 and I2 interfaces forms in the region of the highest defect generation, evidently correlating with the surpassed $\beta$-to-$\gamma$ phase transformation threshold in this region [12]. 
In the neighboring region, between I2 and I1 interfaces, the survival of $\beta$-\ce{Ga2O3} is because of the efficient out-diffusion of defects towards the surface, so that ``dynamically'' the disorder-threshold needed for the irreversible $\beta$-to-$\gamma$ polymorph conversion is never reached in this region. 
Concurrently, this process supplies defects to the very near surface region above I1 interface, so that a thin top $\gamma$-layer forms. 
In contrast, defect diffusion into the bulk may be less efficient than towards the surface explaining the higher residual disorder level behind I3 interface, therefore, resulting in the formation of the mixed $\beta+\gamma$ layer in this region. 

\begin{figure*}[ht!]
 \includegraphics[width=14cm]{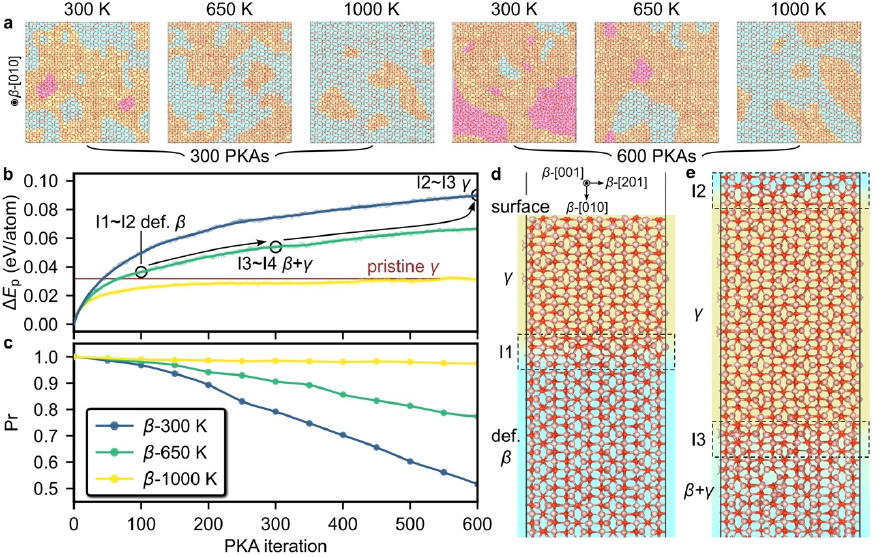}
 \caption{
    \textbf{ML-MD simulation data supporting interpretations of the $\beta$-to-$\gamma$ transformation controlled by dynamic defect annealing.} 
    (a) Final cell-states viewed from $\beta$-\hkl[010] direction, after exposure to 300/600 PKAs as a function of temperature. The pink, yellow, and cyan regions indicate the transiently disordered, $\gamma$-like, and defective $\beta$ phases, respectively. 
    (b) Evolution of the potential energy ($\Delta E_\mathrm{p}$) in the cells. The brown horizontal line is the pristine $\gamma$ phase energy baseline. 
    (c) The Pearson correlation coefficients, Pr, calculated within the 2$^\mathrm{nd}$ shell for the PRDF of the increasingly damaged $\beta$-Ga sublattice with respect to the pristine $\beta$-Ga sublattice PRDFs as a function of the PKA iteration. 
    (d, e) Atomic configurations of the polymorph interfaces corresponding to obtained from ML-MD simulations and labeled I1-I3 in accordance with that found in the experiments.
 }
 \label{fig:4}
\end{figure*}

This phenomenological scenario is consistent with the results from the ML-MD simulations. 
Specifically, Fig.~4 illustrates the structural evolutions of the overlapping cascades with primary knock-on atoms (PKAs) as a function of temperature in $\beta$-cells, extending the methodology developed earlier for RT irradiations [10,12]. 
Evidently, variations in temperature – which essentially means variations in defect annihilation rates – lead to distinct differences in the final disorder levels in the cells. 
Indeed, as seen from Fig.~4(a), at RT the final cell state contains significant proportions of transiently disordered (pink) and $\gamma$-like (yellow) regions, favoring a swift $\beta$-to-$\gamma$ transition [10,12]. 
However, at elevated temperatures the transiently disordered region eventually vanishes, resulting in an increased fraction of the defective $\beta$ (cyan) region, containing point defects. 
This trend agrees with the evolution of the potential energy ($\Delta E_\mathrm{p}$) and Pearson correlation analysis (Pr) of the Ga-Ga partial radial distribution functions (PRDFs), as illustrated in Figs.~4(b) and (c); find additional explanations in Supplementary Note VII. 
The less steep slopes of the $\Delta E_\mathrm{p}$ elevated-temperature curves in Figs.~4(b), as well as decreasing trends in Pr curves in Fig.~4(c), indicate significantly reduced disorder accumulation and less pronounced structural changes during the overlapping cascades, as compared to that at RT. 
For example, following $\Delta E_\mathrm{p}$ analysis, the irradiation performed at 1000 K is unlikely to lead towards $\beta$-to-$\gamma$ transformation, since the corresponding $\Delta E_\mathrm{p}$ curve (yellow colored in Fig.~4(b)) remains below the $\gamma$-pristine energy level. 
On contrary, simulations of 650 K irradiation – consistently with experimentally observed multilayered polymorph self-assembling – reveal options for $\beta$-to-$\gamma$ transformations as seen from the data in Fig.~4(b). 
Indeed, experimentally discovered I1-I3 interfaces in Fig.~1(c), closely correlate with appropriate ML-MD disorder levels data as labeled in Fig.~4(b). 
Furthermore, Figs.~4(d) and (e) illustrate the formation of vertically aligned \hkl(010)$_{\beta}$ $\vert \vert$ \hkl(-110)$_{\gamma}$ interfaces obtained in simulations, nicely matching the interface structure observed experimentally in Fig.~2.

In conclusion, using the disorder-induced ordering approach we demonstrated the fabrication of multilayered \ce{Ga2O3} polymorph structures, challenging to obtain otherwise via conventional ways of synthesis. 
The rationale behind our approach was in finding an appropriate dynamic annealing regime, determined by the irradiation temperature and ion flux, allowing the self-assembling of several $\gamma$/$\beta$ polymorph interfaces, instead of the continuous $\gamma$-film formation on top of the $\beta$-substrate as a result of the disorder-induced ordering at room temperature, as demonstrated earlier in literature [7-12]. 
Tackling this idea, we performed a systematic investigation, combining experimental techniques with simulations, and demonstrated \ce{Ga2O3} structures with two $\gamma$/$\beta$ interface repetitions obtained in one ion implantation step. 
Importantly, the material adjusted to the interfaces conserved the single crystal structure. 
Furthermore, it exhibited repeatable crystallographic stackings maintained at the interfaces and performed differently optically. 
As such, the finding made in this paper paves the way for fabrication of multilayered \ce{Ga2O3} polymorphs with ion beams, potentially emerging as a suitable technology for polymorph heterostructure fabrication.  

\section*{Methods}

Single crystalline \hkl(010) oriented $\beta$-\ce{Ga2O3} wafers purchased from Novel Crystal Technology Inc. were irradiated with 400 keV $^{58}$Ni$^{+}$ ions keeping the ion fluence at a constant value of $1\times10^{16}$ cm$^{-2}$ known to be sufficient for the $\beta$-to-$\gamma$ transition to occur at room temperature (RT) [7]. 
On the other hand we varied the irradiation temperature in the range from RT to 400 $\degree$C keeping the ion flux constant at $6\times10^{12}$ at./(cm$^{2}$ s) for this set of samples. 
A control sample was fabricated using the lower ion flux of $2\times10^{12}$ at./(cm$^{2}$ s). 
All bombardments were performed at 7$\degree$ off-angle orientation from normal direction to minimize channeling.
Notably, even though Ni may potentially chemically interact with the \ce{Ga2O3} matrix, it is known from literature \hkl[10] that at the concentrations used in this work, Ni is predominantly situated in the interstitial positions and does not make any chemical impact on the polymorph transformation either. 

After irradiations the samples were analyzed by a combination of Rutherford backscattering spectrometry in channeling mode (RBS/C), x-ray diffraction (XRD) and analytical transmission electron microscopy (TEM) and scanning Fourier transform infrared nano-spectroscopy (nano-FTIR). 
RBS/C measurements were performed using 1.6 or 2.5 MeV He$^{+}$ ions incident along \hkl[010] direction and backscattered into a detector placed at 165$\degree$ relative to the incident beam direction. 
It should be noted, that in the present paper we limit our consideration to the Ga-sublattice disorder, because of higher sensitivity of the RBS technique to heavy elements as compared to light atoms. 
The 2$\Theta$ XRD measurements were performed using the Bruker AXS D8 Discover diffractometer with high-resolution Cu K$_{\alpha1}$ radiation selected by a triple-bounce Ge \hkl(022) asymmetric monochromator. 

A cross-section lamella has been prepared for TEM studies by focused ion beam. 
TEM investigations were performed using a Tescan Lyra III XMU instrument. 
Analytical TEM and high-resolution TEM (HRTEM) measurements were performed with JEM ARM200F and JEM 2100 transmission electron microscopes operated at 200 kV. 
Elemental analysis by Energy Dispersive X-ray Spectroscopy (EDS) was applied to map the spatial distribution of the implanted Ni ions into the \ce{Ga2O3} matrix.
The near-field optical signatures of the polymorphs and their structural arrangements on a nanoscale were analyzed by nano-FTIR. 
Near-field nano-FTIR measurements were carried out at room temperature using a commercial \textit{IR}-neaSCOPE$^\mathrm{+s}$ system (attocube systems AG) equipped with a broadband mid-infrared laser source and providing chemical analysis and field mapping at 10 nm spatial resolution. 
The nanoscale tip-enhanced IR absorption (reflection) spectra, represented by the second harmonic of the imaginary part (amplitude) of the complex spectrum, were acquired in the spectral range 620-1400 cm$^{-1}$ with resolution of 3 cm$^{-1}$ and 16 cm$^{-1}$. 

The machine-learned molecular dynamics (ML-MD) simulations were conducted using LAMMPS package [22]. 
The self-developed ML interatomic potential of \ce{Ga2O3} system was employed [23], which was designed with the high accuracy for all five experimentally known \ce{Ga2O3} polymorphs and generality for disordered structures. 
The overlapping cascade simulations are conducted using $\beta$-\ce{Ga2O3} cells comprising 81,920 atoms in a cell with the side lengths of $\sim 97 \times 99 \times 95$ \r A$^{3}$. 
The cells are initially thermalized at the preset temperatures of 300/650/1000 K. 
During each cascade, a Ga or O atom is randomly selected as the PKA. 
The kinetic energy of the PKA is set to 1.5 keV with a uniformly random momentum direction. 
The collision cascade process is simulated in the microcanonical ensemble for 5,000 MD steps utilizing a conventional adaptive MD step algorithm to guarantee numerical stability. 
Additionally, electron-stopping frictional forces are applied to the atoms possessing kinetic energies above 10 eV [24]. 
After the cascade process, the simulations proceed further in the quasi-canonical ensemble with Langevin thermostat [25] applied to border atoms for 10 ps at 300/650/1000 K. 
Post processing and visualization of structural data are made possible by the OVITO [26] software. 
For a quantitative comparison of the degree of correlation (similarity) between any two given curves, we utilize the Pearson correlation coefficient, Pr, calculated with the formula:
\begin{equation}
 \mathrm{Pr} = \frac{\sum_{i=1}^{n} (A_{i}-\overline{A})(B_{i}-\overline{B})}{\sqrt{[\sum_{i=1}^{n} (A_{i}-\overline{A})^{2}][\sum_{i=1}^{n} (B_{i}-\overline{B})^{2}]}},
\end{equation}
where $A_{i}$ and $B_{i}$ are the variable samples of the two curves, respectively, and $\overline{A}$ and $\overline{B}$ are the mean values of the variable samples, respectively.
Specifically, Pr is calculated within the 2$^\mathrm{nd}$ shell for the Ga-Ga PRDF (Supplementary Note VII), which has been demonstrated as a sensitive characterization parameter of the structural changes in $\beta$-\ce{Ga2O3} system in our previous works [10,12]. 
$\mathrm{Pr} \simeq 1$ reflects a positive correlation (similarity) between the curves, whereas $\mathrm{Pr} \simeq 0$ and $\mathrm{Pr} \simeq -1$ indicate no correlation (dissimilarity) and a negative correlation (anti-similarity), respectively. The PRDF is defined as:
\begin{equation}
   g_\mathrm{p}^\mathrm{XY}(r) = [ \frac{1}{c_\mathrm{X}N_\mathrm{at.}}\sum_{i=1}^{c_\mathrm{X}N_\mathrm{at.}}\frac{n_{i}^\mathrm{XY}(r)}{4\pi r^{2}dr}]/(c_\mathrm{Y}N_\mathrm{at.}/V), 
\end{equation}
where $\mathrm{X}$ and $\mathrm{Y}$ stand for the element species of the central and surrounding atoms, respectively; $c_\mathrm{X}$ and $c_\mathrm{Y}$ are the corresponding atomic concentrations of $\mathrm{X}$ and $\mathrm{Y}$ in the system.

\section*{Acknowledgments}

M-ERA.NET Program is acknowledged for financial support via GOFIB project (administrated by the Research Council of Norway project number 337627). AA and EM acknowledge the Research Centre for Sustainable Solar Cell Technology (FME SuSolTech, RCN project number 257639). The Research Council of Norway is acknowledged for the support to the Norwegian Micro- and Nano-Fabrication Facility, NorFab, project number 295864. C.R., I.F.M. and C.G. acknowledge funding from project PC2-PN23080202 within the Core Program of PNCDI 2022-2027. J.Z. acknowledges the National Natural Science Foundation of China under Grant 62304097; Guangdong Basic and Applied Basic Research Foundation under Grant 2023A1515012048; Shenzhen Fundamental Research Program under Grant JCYJ20230807093609019. The computational resources are provided by the Center for Computational Science and Engineering at the Southern University of Science and Technology and the IT Center for Science, CSC, Finland. The international collaboration was in part enabled by the INTPART program at the Research Council of Norway via project number 322382.

\section*{Author contribution}

A.K. and A.A. conceptualized the work; A.A. and V.V. collected the channeling and x-ray diffraction data, respectively, C.R., I.F.M., and C.G. collected the electron microscopy data and performed corresponding HRTEM simulations; A.G. and A.C. collected nano-FTIR data. A.K. and A.A. formulated the phenomenological model. J.Z. and F.D. conducted the ML-MD simulations and analyzed of the computational data. A.A., C.G., and A.K. composed the initial draft of the manuscript. All coauthors read and commented on the initial draft. A.K., A.A., J.Z., and C.G. finalized the manuscript with inputs from all co-authors. All co-authors discussed the results as well as reviewed and approved the manuscript. E.M., C.G., J.Z., F.D., and A.K. administrated their parts of the project and contributed to the funding acquisition.

\section*{Competing interests} 

The authors declare no competing interests.

\section*{References}

\noindent$^{1}$ Kroemer, H. Nobel Lecture: Quasielectric fields and band offsets: teaching electrons new tricks, \textit{Rev. Mod. Phys.} \textbf{73}, 783 (2001).

\noindent$^{2}$ Gentili, D., Gazzano, M., Melucci, M., Jones, D. \& Cavallini, M. Polymorphism as an additional functionality of materials for technological applications at surfaces and interfaces. \textit{Chem. Soc. Rev.} \textbf{48}, 2502 (2019).

\noindent$^{3}$ Sung, J., Heo, H., Si, S. et al. Coplanar semiconductor–metal circuitry defined on few-layer \ce{MoTe2} via polymorphic heteroepitaxy. \textit{Nat. Nanotechol.} \textbf{12}, 1064–1070 (2017). 

\noindent$^{4}$ Tak, B. R. et al. Recent advances in the growth of gallium oxide thin films employing various growth techniques - a review. \textit{J. Phys. D: Appl. Phys.} \textbf{54}, 453002 (2021).

\noindent$^{5}$ Biswas, M. \& Nishinaka, H. Thermodynamically metastable $\alpha$-, $\epsilon$- (or $\kappa$-), and $\gamma$-\ce{Ga2O3}: From material growth to device applications. \textit{APL Mater.} \textbf{10}, 060701 (2022).

\noindent$^{6}$ Lee, J., Gautam et al. Investigation of enhanced heteroepitaxy and electrical properties in $\kappa$-\ce{Ga2O3} due to interfacing with $\beta$-\ce{Ga2O3} template layers. \textit{Phys. Status Solidi A} \textbf{220}, 2200559 (2023).

\noindent$^{7}$ Azarov, A. et al. Disorder-induced ordering in gallium oxide polymorphs. \textit{Phys. Rev. Lett.} \textbf{128}, 015704 (2022).

\noindent$^{8}$ Anber, E. A. et al. Structural transition and recovery of Ge implanted $\beta$-\ce{Ga2O3}. Appl. Phys. Lett. 117, 152101 (2020).

\noindent$^{9}$ Yoo, T. et al. Atomic-scale characterization of structural damage and recovery in Sn ion-implanted $\beta$-\ce{Ga2O3}. \textit{Appl. Phys. Lett.} \textbf{121}, 072111 (2022).

\noindent$^{10}$ Azarov, A. et al. Universal radiation tolerant semiconductor. \textit{Nat. Commun.} \textbf{14}, 4855 (2023). 

\noindent$^{11}$ Huang H.-L., Chae C., Johnson J. M., Senckowski A., Sharma S., Singisetti U., Wong M. H., and Hwang J. Atomic scale defect formation and phase transformation in Si implanted $\beta$-\ce{Ga2O3}. \textit{APL Mater.} \textbf{11}, 061113 (2023).

\noindent$^{12}$ Zhao, J., Garc{\'i}a-Fern{\'a}ndez, J., Azarov, A., He, R., Prytz, {\O}., Nordlund, K., Djurabekova, F., Hua, M. \& Kuznetsov, A. Crystallization instead amorphization in collision cascades in gallium oxide. \textit{arXiv} DOI: 10.48550/arXiv.2401.07675 (2024)

\noindent$^{13}$ Peres, M. et al. Doping $\beta$-\ce{Ga2O3} with europium: influence of the implantation and annealing temperature. \textit{J. Phys. D: Appl. Phys.} \textbf{50}, 32510 (2017).

\noindent$^{14}$ Azarov, A., Venkatachalapathy, V., Monakhov, E. V. \& Kuznetsov, A. Yu. Dominating migration barrier for intrinsic defects in gallium oxide: dose-rate effect measurements. \textit{Appl. Phys. Lett.} \textbf{118}, 232101 (2021).

\noindent$^{15}$ Pearton, S. J. et al. A review of \ce{Ga2O3} materials, processing, and devices. \textit{Appl. Phys. Rev.} \textbf{5}, 011301 (2018).

\noindent$^{16}$ Playford, H. Y., Hannon, A. C., Barney, E. R. \& Walton, R. I. Structures of uncharacterised polymorphs of gallium oxide from total neutron diffraction. \textit{Chem. Eur. J.} \textbf{19}, 2803 – 2813 (2013)

\noindent$^{17}$ Playford, H. Y. et al. Characterization of structural disorder in $\gamma$‑\ce{Ga2O3}. \textit{J. Phys. Chem. C} \textbf{118}, 16188-16198 (2014)

\noindent$^{18}$ H{\"y}tch, M. J., Snoeck, E., Kilaas, R. Quantitative measurement of displacement and strain fields from HREM micrographs. \textit{Ultramicroscopy} \textbf{74}, 131-146 (1998).

\noindent$^{19}$ GPA for DigitalMicrograph, HREM Research Inc., http://www.hremresearch.com/ 

\noindent$^{20}$ Seto Y. \& Ohtsuka, M. ReciPro: free and open-source multipurpose crystallographic software integrating a crystal model database and viewer, diffraction and microscopy simulators, and diffraction data analysis tools. \textit{J. Appl. Crystallogr.} \textbf{55}, 397-410 (2022).

\noindent$^{21}$ Ziegler, J. F., Ziegler, M. D. \& Biersack, J. P. SRIM—the stopping and range of ions in matter (2010). \textit{Nucl. Instrum. Methods Phys. Res. B} \textbf{268}, 1818 (2010).

\noindent$^{22}$ Plimpton S., Fast parallel algorithms for short-range molecular dynamics. \textit{J. Comp. Phys.} \textbf{117}, 1 (1995).

\noindent$^{23}$ Zhao J., Byggm{\"a}star J., He H., Nordlund K., Djurabekova F., \& Hua M., Complex \ce{Ga2O3} polymorphs explored by accurate and general-purpose machine-learning interatomic potentials. \textit{npj Comput. Mater.} \textbf{9}, 159 (2023).

\noindent$^{24}$ Nordlund K., Ghaly M., Averback R. S., Caturla M., Diaz de la Rubia T., \& Tarus J., Defect production in collision cascades in elemental semiconductors and fcc metals. \textit{Phys. Rev. B} \textbf{57}, 7556 (1998).

\noindent$^{25}$ D{\"u}nweg B. \& Paul W., Brownian dynamics simulations without Gaussian random numbers. \textit{Int. J. Mod. Phys. C} \textbf{02}, 817 (1991).

\noindent$^{26}$ Stukowski A., Visualization and analysis of atomistic simulation data with OVITO–the open visualization tool. \textit{Modell. Simul. Mater. Sci. Eng.} \textbf{18}, 015012 (2010).




\end{document}